# Test of T and CP Violation in Leptonic Decay of $\tau^{\pm\,*}$

YUNG SU TSAI

*Stanford Linear Accelerator Center*
*Stanford University, Stanford, California 94309*

## ABSTRACT

The $\tau^{\pm}$, highly polarized in the direction of the incident beam, can be obtained from the $e^{\pm}$ collider with the polarized incident $e^-$ (and preferably also the $e^+$) beam. This polarization vector $\vec{w}_i = (w_1 + w_2)/(1 + w_1 w_2)\hat{e}_z$ can be used to construct the $T$ odd rotationally invariant product $(\vec{w}_i \times \vec{p}_\mu) \cdot \vec{w}_\mu$, where $w_1$ and $w_2$ are longitudinal polarization vectors of $e^-$ and $e^+$ respectively; $\vec{p}_\mu$ and $\vec{w}_\mu$ are the momentum and polarization of the muon in the decay $\tau^- \to \mu^- + \overline{\nu}_\mu + \nu_\tau$. $T$ is violated by the existence of such a term. $CP$ can be tested by comparing it with a similar term in $\tau^+$ decay. If $T$ violation in such a decay is milliweak or stronger, one can find it using the proposed polarized $\tau$-charm factory with luminosity of $1 \sim 3 \times 10^{33}/\text{cm}^2/\text{sec}$. One can test whether $T$ (and $CP$) violation is due to the charged Higgs boson exchange by doing a similar experiment for the $\mu^{\pm}$ decay.

To be presented at the "Workshop on the Tau/Charm Factory,"
21–23 June 1995, Argonne National Laboratory

Submitted to Physical Review Letters.

___

*Work supported by the Department of Energy, contract DE–AC03–76SF00515.

In the Standard Model of Kobayashi and Maskawa [1] $CP$ violation occurs as a result of a complex phase in the unitary matrix relating gauge eigenstates and mass eigenstates. The leptonic sector does not have $CP$ violation if all neutrinos are massless. Both of these assumptions could be wrong. It is quite possible [2] that $CP$ violation is due to exchange of some new particle such as a heavier $W$ boson or a charged Higgs boson. If $CP$ violation is milliweak or stronger in $\tau$ decay, one should be able to observe it in the proposed $\tau$-charm factory where it is expected to have $1 \sim 3 \times 10^8$ highly polarized $\tau$ pairs per year [2].

The $CP$ violation in $\tau$ production can be ignored because we are dealing with electromagnetic production. The radiative correction due to $CP$ violation in the weak interaction is of order $10^{-5}$ if it is weak, but $10^{-8}$ if it is semiweak [3]. In contrast to the production, the decay of $\tau$ is weak, thus $CP$ violation is of order 1 if it is weak and $10^{-3}$ if it is milliweak. Up to now, the only $CP$ violation is from $K_L$ which is $2 \times 10^{-3}$. In a recent paper [2] we dealt with the $CP$ violation in the semileptonic decay of $\tau$ with two or more final hadrons. For a single hadron in the semileptonic decay or a leptonic decay, the only rotationally invariant quantity we can form is $\vec{w}_i \cdot \vec{q}$ where $\vec{q}$ is the momentum of the final visible particle. But this term is $T$ even so we cannot have $CP$ violating effects from this term without violating TCP [4]. It is very desirable to measure $CP$ violation in pure leptonic decay because in the semileptonic decay it is impossible to assign $CP$ violation to the leptonic or hadronic vertex [2]. For the leptonic decay we have to measure the polarization of the muon and construct a rotationally invariant product

$$(\vec{w}_i \times \vec{p}_\mu) \cdot \vec{w}_\mu \ , \tag{1}$$

where

$$\vec{w}_i = \frac{w_1 + w_2}{1 + w_1 w_2} \hat{e}_z \ , \tag{2}$$

with $w_1$ and $w_2$ being the longitudinal polarization of the incident electron and positron respectively; $\vec{p}_\mu$ is the laboratory momentum, and $\vec{w}_\mu$ the polarization of the muon. The muon polarization is measured by the asymmetry in electron distribution coming from the term $\vec{w}_\mu \cdot \vec{q}_{e^-}$ where $\vec{q}_{e^-}$ is the electron momentum. Thus the correlation in Eq. (1) induces the correlation

$$c(\vec{w}_i \times \vec{p}_\mu) \cdot \vec{q}_{e^-} \ , \tag{3}$$

which is also odd under $T$.

Equation (3) means that if one finds an asymmetry in the perpendicular component of $\vec{q}_{e^-}$ with respect to the plane formed by $\vec{w}_i$ and $\vec{p}_\mu$, one discovers the existence of $T$ violating effect. Under $CP$ we have $w_1 \leftrightarrow w_2$, $w_i \to w_i$, $\vec{p}_{\mu^-} \leftrightarrow -\vec{p}_{\mu^+}$, $\vec{q}_{e^-} \leftrightarrow -\vec{q}_{e^+}$. Thus if $CP$ is conserved, we have for $\tau^+$ decay

$$c'(\vec{w}_i \times \vec{p}_{\mu^+}) \cdot \vec{q}_{e^+} \tag{4}$$

with $c' = c$. But since $T$ is violated by both Eqs. (3) and (4), we better have $c' = -c$ in order to preserve $TCP$ invariance. The discussion given above is completely model independent. Later we shall give a model which will illustrate all the above observations.

In order to measure the angular asymmetry in the decay electrons, we have to slow the muon down to almost at rest. For $\tau$ energy $E_\tau$ equal to 2.087 GeV, where the cross section is maximum [2], the maximum and minimum muon momenta are respectively:

$$p_3^{\max} = \frac{E_\tau}{2}\left[(1+\beta) - \frac{\mu^2}{M^2}(1-\beta)\right] = 1.589 \text{ GeV} \tag{5}$$

$$p_3^{\min} = -\frac{E_\tau}{2}\left[(1-\beta) - \frac{\mu^2}{M^2}(1+\beta)\right] = -0.4904 \text{ GeV} \tag{6}$$

where

$$\mu = 0.105658 \text{ GeV}$$
$$\beta = \sqrt{1.5 - \sqrt{1.5}}$$
$$M = 1.777 \text{ GeV}$$

$p_3^{\min}$ is negative means that it is going in the opposite direction to the $\tau$ momentum. Approximately 10% of muons are going backward. The asymmetry caused by the detector can be checked by reversing the polarization of the incident beam (or beams).

In this paper we use the same model of $T$ and $CP$ violation as the previous paper [2]. It is shown there that if we limit the weak interaction to be transmitted by exchange of spin 1 and spin 0 particles, then we have only two possible choices of matrix elements denoted by $M_1$ and $M_2$ (see Fig. 1) that can interfere with the Standard Model matrix denoted by $M_0$.

$$M_0 = A\,\overline{u}(p_2)\gamma_\mu(1-\gamma_5)u(p_1)\overline{u}(p_3)\gamma_\mu(1-\gamma_5)v(p_4), \tag{7}$$

$$M_1 = B\,\overline{u}(p_2)\gamma_\mu(1-\gamma_5)u(p_1)\overline{u}(p_3)\gamma_\mu(1-\gamma_5)v(p_4), \tag{8}$$

$$M_2 = C\,\overline{u}(p_2)(1+\gamma_5)u(p_1)\overline{u}(p_3)(1-\gamma_5)v(p_4), \tag{9}$$

where $p_1, p_2, p_3$ and $p_4$ are momenta of $\tau^-$, $\nu_\tau$, $\mu^-$ and $\overline{\nu}_\mu$ respectively. We have assumed that $m_{\nu_\mu}/m_\mu$ and $m_{\nu_\tau}/m_\tau$ to be either zero or too small to be experimentally observable, so that possible terms such as $(1+\gamma_5)u(p_4)$ and $(1+\gamma_5)u(p_2)$ are ignored in $M_1$ and $M_2$. $A$ is chosen to be real while $B$ and $C$ are allowed to be complex. Since there is no final state interaction the imaginary parts of $B$ and $C$ cause $T$ violating effects. If $TCP$ is conserved then $B$ and $C$ for the $\tau^+$ decay must be the complex conjugate of $B$ and $C$:

$$\overline{B} = B^* \quad \text{and} \quad \overline{C} = C^* . \tag{10}$$

Since the Standard Model is good to $10^{-3}$ to $10^{-2}$, we can assume $M_1^+ M_1$ and $M_2^+ M_2$ to be at most $10^{-2}$ compared to $M_0^+ M_0$ and thus we shall ignore them. We shall also ignore $M_1$ completely because its interference with $M_0$ does not depend upon the imaginary part of $B$ that causes the $T$ violation:

$$M_0^+ M_1 + M_1^+ M_0 = (B + B^*)M_0^+ M_0/A . \tag{11}$$

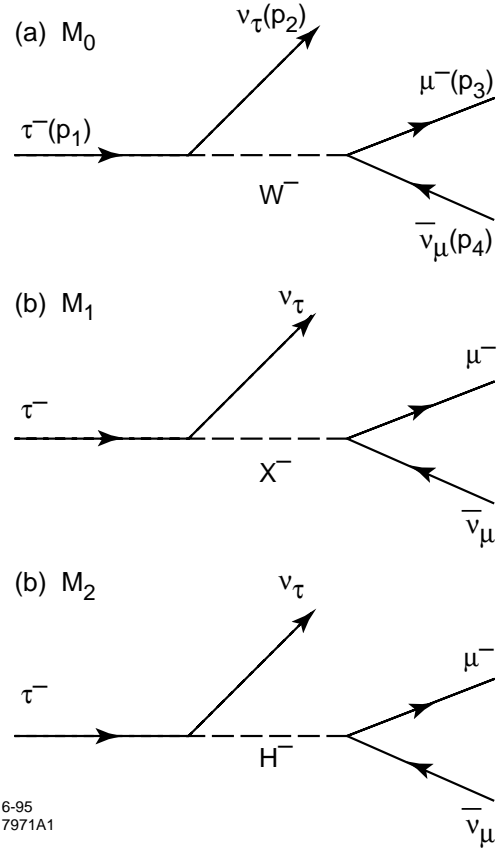

Figure 1: (a) $M_0$: Feynman diagram for $\tau^- \to \mu^- + \overline{\nu}_\mu + \nu_\tau$ in the Standard Model that conserve $T$ and $CP$. (b) $M_1$: A possible $T$ violating spin 1 exchange diagram that is shown not to contribute to the $T$ violating effect. (c) $M_2$: A $T$ violating spin 0 exchange diagram with a complex coupling constant.

Writing $M_0 = A\mathcal{M}_0$ and $M_2 = C\mathcal{M}_2$, we have

$$M_0^+ M_2 + M_2^+ M_0 = A\operatorname{Re} C(\mathcal{M}_0^+ \mathcal{M}_2 + \mathcal{M}_2^+ \mathcal{M}_0) + iA\operatorname{Im} C(\mathcal{M}_0^+ \mathcal{M}_2 - \mathcal{M}_2^+ \mathcal{M}_0). \quad (12)$$

Only the imaginary part of $C$ contributes to $T$ violation. The real part of $B$ and $C$ should be of order $10^{-2}$ or less compared with $A$, thus they will be ignored. We have therefore

$$(M_0 + M_1 + M_2)^+(M_0 + M_1 + M_2) \approx A^2 \mathcal{M}_0^+ \mathcal{M}_0 + iA\operatorname{Im} C(\mathcal{M}_0^+ \mathcal{M}_2 - \mathcal{M}_2^+ \mathcal{M}_0).$$

Let $\vec{w}_3$ be the polarization vector of the muon in the rest frame of the muon. We are interested in the $y$ component of $\vec{w}_3$ defined in Fig. 2(a) whose existence signifies the violation of $T$ because $(\vec{w} \times \vec{p}_3) \cdot \vec{w}_3$ is odd under $T$. After averaging over the $\tau$ production angle the polarization vector of $\tau^-$, $\vec{w}$, is replaced by the initial beam polarization $\vec{w}_i$. We note that since the $y$ direction is perpendicular to $\vec{p}_3$ it is invariant under the Lorentz boost along $\vec{p}_3$. The $y$ component of the muon polarization can be calculated using the formula

$$W_{3y} = \frac{i\operatorname{Im} C \int \left[(\mathcal{M}_0^+ \mathcal{M}_2 - \mathcal{M}_2^+ \mathcal{M}_0)_{\vec{s}=\hat{e}_y} - (\mathcal{M}_0^+ \mathcal{M}_2 - \mathcal{M}_2^+ \mathcal{M}_0)_{\vec{s}=-\hat{e}_y}\right] \times \frac{d^3p_2}{2E_2} \frac{d^3p_4}{2E_4} \delta^4(p_1-p_2-p_3-p_4)}{A \int \sum_{\text{Spin of }\mu} (\mathcal{M}_0^+ \mathcal{M}_0) \frac{d^3p_2}{2E_2} \frac{d^3p_4}{2E_4} \delta^4(p_1 - p_2 - p_3 - p_4)}. \quad (13)$$

The second term inside the square bracket of the numerator is the negative of the first, thus the bracket is equal to twice the first term. The phase space integration with respect to the two undetected neutrons is carried out in the rest frame of $u = p_2 + p_4$. We denote quantities in this frame by $\hat{\ }$.

$$\int \frac{d^3p_2}{2E_2} \frac{d^3p_4}{2E_4} \delta^4(p_1 - p_2 - p_3 - p_4) = \int d\hat{\Omega}_4 \frac{\hat{E}_4}{2} \delta(u^2 - 2u\hat{E}_4) d\hat{E}_4$$

$$= \frac{1}{8} \int d\hat{\Omega}_4, \quad (14)$$

$$(m_0^+ m_2 - m_2^+ m_0)$$
$$= \frac{\operatorname{Tr}}{4}(1+\gamma_5 \slashed{w})(\slashed{p}_1 + M)(1+\gamma_5)\gamma_\mu \slashed{p}_2 \frac{\operatorname{Tr}}{4} \slashed{p}_4(1+\gamma_5)\gamma_\mu(1+\gamma_5 \slashed{s})(\slashed{p}_3+m)$$
$$- \frac{\operatorname{Tr}}{4}(1+\gamma_5 \slashed{w})(\slashed{p}_1+M)(1-\gamma_5)\slashed{p}_2\gamma_\mu \frac{\operatorname{Tr}}{4} \slashed{p}_4(1+\gamma_5)(1+\gamma_5 \slashed{s})(\slashed{p}_3+m)\gamma_\mu$$
$$= 4i\bigg[-(s \cdot p_4)EPS(w,p_1,p_3,p_4) + (p_3 \cdot p_4)EPS(s,w,p_1,p_4)$$
$$+ \frac{1}{2}(p_1 \cdot p_4)EPS(s,w,p_1,p_3) - \frac{1}{2}\{(p_1 \cdot p_3) - m^2\} EPS(s,w,p_1,p_4)\bigg]. \quad (15)$$

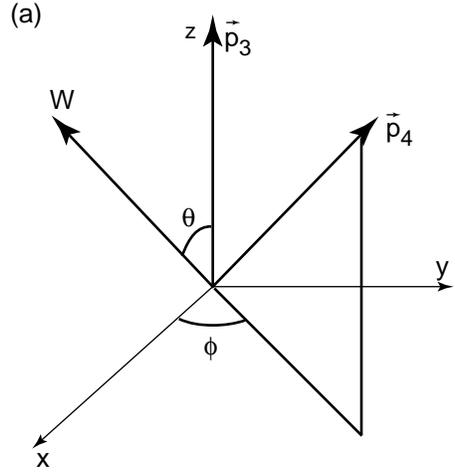

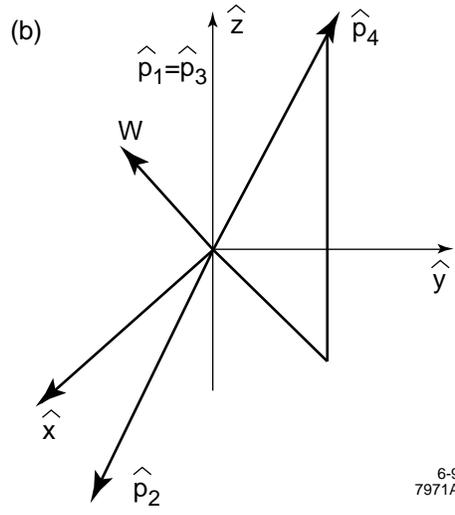

Figure 2: (a) The rest frame of $\tau^-$, the coordinate system used in Eq. (18). (b) The rest frame of $u = p_2 + p_4 = p_1 - p_3$, the coordinate system used in integrating out the two unobserved neutrinos in Eqs. (13) and (14). This frame is obtained from the above diagram by boosting against the direction of $\vec{p}_3$.

In the above we have dropped all those terms that are odd in $\hat{p}_{4x}$, $\hat{p}_{4y}$, and $\hat{p}_{4x}$ because they yield zero after integration with respect to $d\hat{\Omega}_4$. The Levi-Civita EPS's are evaluated in the rest frame of $\tau$ and then $p_{4\mu}$ is Lorentz transformed to $\hat{p}_{4\mu}$ before the angular integration shown in Eq. (14). In the rest frame of $u = p_2 + p_4$ we have $\hat{p}_1 = \hat{p}_3$, thus the Lorenz boost is in the z direction [5]: $E_4 = \gamma \hat{E}_4 - \beta\gamma\hat{p}_{4z}$, $p_{4x} = \hat{p}_{4x}$, $p_{4y} = \hat{p}_{4y}$ and $p_{4z} = -\gamma\beta\hat{E}_4 + \gamma\hat{p}_{4z}$ with $\gamma = (M - E_3)/u$, $\beta = p_3/u$, $u = \sqrt{M^2 + m^2 - 2ME_3}$. The result of the angular integration is

$$\int (\mathcal{M}_0^+ \mathcal{M}_2 - \mathcal{M}_2^+ \mathcal{M}_0)_{\vec{s}=\hat{e}_y} d\hat{\Omega}_4 = \frac{4\pi i}{3} \left[3M^2 - 4E_3 M + m^2\right] EPS(\hat{e}_y, w, p_1, p_3) , \quad (16)$$

with

$$EPS\,(\hat{e}_y, w, p_1, p_3) = M(\vec{w} \times \vec{p}_3)_y .$$

The denominator in Eq. (13) can be obtained similarly:

$$\int \sum_{\text{spin of}\mu} (\mathcal{M}_0^+ \mathcal{M}_0) d\hat{\Omega}_4 = \frac{32\pi\, M^2 E_3}{3} \left[3M - 4E_3 - \frac{2m^2}{E_3} + \frac{3m^2}{M} + (\vec{w}\cdot\vec{p}_3)\left(\frac{M}{E_3} - 4 + \frac{3m^2}{E_3 M}\right)\right] . \quad (17)$$

Equation (17) agrees with the result of my previous paper [6] written several years before the discovery of the $\tau$.

Putting everything together we have finally:

$$W_{3y} = \frac{-(\vec{w} \times \vec{p}_3)_y}{8E_3} \frac{\left[3M - 4E_3 + \frac{m^2}{M}\right] \text{Im}\,(C/A)}{3M - 4E_3 - \frac{2m^2}{E_3} + \frac{3m^2}{M} + (\vec{w}\cdot\vec{p}_3)\left(\frac{M}{E_3} - 4 + \frac{3m^2}{ME_3}\right)} . \quad (18)$$

For $\tau^+$ decay we use the substitution $\vec{p}_3 \to -\vec{p}'_3$, $\vec{w} \to \vec{w}'$, $E_3 \to E'_3$, $C \to \overline{C}$. $\overline{C}$ is equal to $C^*$ if TCP is conserved [2]. Under $CP$ we have $w \to w'$, $\vec{p}_3 = -\vec{p}'_3$, thus it is opposite to the TCP conserved case. Thus $CP$ must be violated in order to conserve TCP.

The polarization of the muon is analyzed by the decay electron momentum $\vec{q}$. Thus the measurement of the existence of the $T$ violating term $(\vec{w} \times \vec{p}_3) \cdot \vec{w}_3$ can be done by measuring the existence of the $T$ violating correlation $(\vec{w} \times \vec{p}_3) \cdot \vec{q}$, where $\vec{p}_3$ and $\vec{q}$ are momenta of muon and decaying electron in the rest frame of $\tau$. Exactly at threshold such a correlation can be calculated using Eq. (18), but as energy is increased one must integrate over the $\tau$ production angle. The result must be proportional to the only $T$ noninvariant correlation in the center-of-mass system $(\vec{w}_i \times \vec{p}_\mu) \cdot \vec{q}_e$, where $\vec{w}_i$ is the initial beam polarization defined in Eq. (2); $\vec{p}_\mu$ and $\vec{q}_e$ are center-of-mass momenta of the muon and electron respectively.

We have shown above that only the spin 0 exchange can produce $T$ violating leptonic decay of the $\tau$. By measuring a similar effect in $\mu^\pm \to e^\pm + \nu_e + \nu_\mu$ one should be able to decipher if the exchanged particle is the Higgs boson discussed by T. D. Lee [7] and S. Weinberg [8]. The test of $T$, $CP$, and charged Higgs boson exchange in the leptonic decay of $\tau$ proposed in this paper as well as the test of $CP$, $TCP$ and $CVC$ in the semileptonic decay of $\tau$ proposed in my previous paper [2] are mostly for the proposed tau-charm factory. However they can also be carried out in the $B$ factories being constructed at SLAC, KEK and Cornell provided they add a capability to longitudinally polarize their initial electron (and preferably also positron) beam. It is regrettable that none of the $B$ factories mentioned above have any plans to polarize their incident beam (or beams). At the $B$ factory energy the cross section is about 1/6 that of the tau-charm factory for producing $\tau$ pairs and the polarization of produced $\tau$ is about 23% less favorable due to the reduced $s$ wave dominance in the production [see Eq. (4.11) of Ref. 2]. However the luminosity of the machine is supposed to be roughly proportional to the energy that is a factor of three in favor of the $B$ factory. Thus the tests proposed in this paper and Ref. [2] are still do-able with the $B$ factories if they polarize their incident beams.

## Acknowledgments

The author wishes to thank Professor W.K.H. Panofsky for reviving my interest in the Tau-Charm Factory. I also would like to thank Bill Dunwoodie, Charles Prescott and Francesco Villa for consultation on measurement of transverse polarization of the muon and electron.

# References

[1] M. Kobayashi and T. Maskawa, *Prog. Theor. Phys.* **49,** 652, (1973).

[2] Y. S. Tsai, *Phys. Rev.* **D51,** 3172 (1995).

[3] The effect due to the electric dipole moment of $\tau$ can be regarded as the weak or milliweak radiative corrections to the electromagnetic vertex of $\tau$ and thus the correction must be of order $(m_\tau/m_w)^2 \alpha = 3 \times 10^{-6}$ if it is weak and another factor of $10^{-3}$ if it is milliweak. The interference of one $\gamma$ exchange and one Higgs boson exchange is $(m_e/E)(E^2/m_H^2) \, m_\tau m_e/m_w^2 \approx (m_e^2 m_\tau^2)/(m_H^2 m_w^2) = 1.23 \times 10^{-10} \text{GeV}^2/m_H^2$.

[4] T. D. Lee made this valuable remark.

[5] For readers interested in the technical aspects of the calculation, the choice of the rest frame of $u = (p_1 - p_3) = (p_2 + p_4)$ to do the angular integration, the choice of the direction of $\vec{p}_3$ as the $z$ axis to facilities Lorenz transformation and the choice of the rest frame of $\tau$ to express the final result are very important in expediting the whole calculation.

[6] Y. S. Tsai, *Phys. Rev.* **4D,** 2821 (1971). See Eq. (2.1′).

[7] T. D. Lee, *Phys. Rev.* **D8,** 1226 (1973); Phys. Rep. **9C,** 143 (1974).

[8] S. Weinberg, *Phys. Rev. Lett.* **37,** 657 (1976).